\documentclass[aps,prc,showpacs,twocolumn,preprintnumbers,floatfix, showkeys,tightenlines]{revtex4}
\usepackage{amsmath,amssymb,amsfonts}

\usepackage{amsmath,amssymb,amsfonts}
\usepackage{graphicx}
\usepackage{color}
\usepackage{floatflt}
 \usepackage{longtable}
 \usepackage{dcolumn}
 \usepackage{bm}
\usepackage{color}

\allowdisplaybreaks

\newcommand{\be}{\begin{equation}}
\newcommand{\ee}{\end{equation}}
\newcommand{\bea}{\begin{eqnarray}}
\newcommand{\eea}{\end{eqnarray}}

\begin{document}



\title{ Diversities in the properties of neutron stars 
at a \\fixed neutron-skin thickness in $^{208}$Pb nucleus}
\author{N. Alam$^1$}
\author{A. Sulaksono$^2$ }
\author{B. K. Agrawal$^1$}
\email{bijay.agrawal@saha.ac.in}

\affiliation{ $^1$ Saha Institute of Nuclear Physics, 1/AF Bidhannagar, Kolkata {\sl 700064}, India \\
$^2$ Departemen Fisika, FMIPA, Universitas Indonesia, Depok, 16424, Indonesia.}


\begin{abstract}

We study the diversities in the properties of the neutron stars arising
due to the different choices for the cross-coupling between various mesons
which governs the density dependence of the nuclear symmetry energy in the
extended relativistic mean-field(RMF) model. For this purpose, we obtain
two  different families of the extended RMF model corresponding to
different non-linear cross-coupling term in the isovector part of the
effective Lagrangian density. The lowest order contributions for the
$\delta$ mesons are also included. The different models within the same family
are so obtained that they yield wide variation in the value of neutron-skin
thickness in the $^{208}$Pb nucleus. These models are employed to compute
the neutron star properties such as, core-crust transition density, radius
and red shift at canonical mass ($1.4M_{\odot}$), tidal polarizability
parameter, and threshold mass required for the enhanced cooling through
direct Urca process. Most of the neutron star properties
considered are significantly different(10\%-40\%) for the different
families of  models at a smaller neutron-skin thickness ($\sim 0.15$ fm) in the
$^{208}$Pb nucleus.

\end{abstract}


\maketitle
\newpage
\section{Introduction}

The neutron stars are believed to be composed of highly asymmetric matter,
predominantly, the  neutrons. Of course, a small admixture of protons,
electrons and muons are also present   to maintain the $\beta$-equilibrium
and charge neutrality. The various theoretical models have conjectured the
possibility of existence of exotica, like, hyperons, Bose condensates and
quarks in the core of the neutron stars
\cite{Glendenning91,Glendenning98,Lackey06,Schulze06,Lattimer07,Bednarek07,Schaffner00,Chamel13}. 
The precise knowledge of the masses and the radii of the neutron stars are
crucial in determining their compositions.  Recently, significant progress
has been made along this direction \cite{Lattimer12}. The masses of PSR
J1614-2230 \cite{Demorest10} and PSR J0348+0432 \cite{Antoniadis13}
are measured to be $1.97\pm 0.04M_\odot$ and $2.01\pm 0.04M_\odot$,
respectively.  These measurements impose  the lower bounds on the
maximum mass of the neutron stars that a theoretical model must
yield. In the absence of the exotic degrees of freedoms, the criterion
of maximum mass to be $\sim 2M_\odot$ is readily satisfied by the
theoretical models. The observational constraints on the maximum mass
of neutron stars do not completely rule out the possibility of existence
of the exotic degrees of freedom,
 but,  the threshold transition densities for their appearance
are pushed to 2.5-3.5 times the nuclear saturation
density\cite{Lonardoni15,Lim14,Sulaksono12}.  The radii of neutron stars
are known only poorly.  The  values of radii are quite sensitive to the
assumed composition of the atmosphere.  The radius $R_{1.4} = 10.7 -
13.1$km, for the neutron star with  the canonical mass of $1.4M_\odot$,
is found to be consistent with the observational analysis and the host
of experimental data for finite nuclei \cite{Lattimer13}.

The equations of state (EOSs) for the $\beta$-equilibrated baryonic
matter, employed to determine the bulk properties of neutron stars,
are usually constructed using the energy density functionals derived from the
Skyrme type effective forces or from an effective Lagrangian density
associated  with the relativistic mean field (RMF) model.
Often, the energy density functionals are optimized  using some selected
experimental data on a key properties of the finite nuclei. Occasionally,
pseudo-data on nuclear and neutron matter are also used in the optimization
protocols \cite{Chabanat97,Agrawal12,Fattoyev13a}. The EOS for the
nuclear matter at a given  density and the asymmetry can be viewed for
simplicity as,
\begin{equation}
\epsilon(\rho,I)=\epsilon(\rho,0)+S(\rho)I^2
\end{equation}
where, $\epsilon(\rho,I)$ is the  energy per nucleon, $\rho =
\rho_n+\rho_p$, $I = (\rho_n - \rho_p)/\rho$ is the asymmetry
with $\rho_n$ and $\rho_p$ being densities for the neutrons and the
protons. The $\epsilon(\rho,0)$   is the EOS for the symmetric  nuclear
matter and $S(\rho)$ is the symmetry energy coefficient at a density
$\rho$.  The EOS of the symmetric nuclear matter for the densities up to $4.5$ times saturation density
($\rho_0=0.16\text{ fm}^{-3}$)
is derived within the
reasonable limits by combining the experimental data
for the finite nuclei with the collective flow and kaon production data
in heavy-ion collisions \cite{Danielewicz02,Fuchs06,Fantina14}. 
The poorly known density dependence of $S(\rho)$ is the major source
for  uncertainty in the EOS for the asymmetric matter .  The value of
$S(\rho)$ is reasonably constrained only around the saturation density 
by the bulk properties of the finite nuclei.
The understanding of density dependence of the $S(\rho)$ is crucial
as it controls the radii of neutron stars, the thicknesses of their
crusts, the rate of cooling of neutron stars, and the properties of
nuclei involved in r-process nucleosynthesis.  The density dependence
of the symmetry energy around the saturation density appears to be well
correlated with the neutron-skin thickness $\Delta r_{\rm np}$ in a heavy
nucleus which can be experimentally measured.  The $\Delta r_{\rm np}$
is the difference between the rms radii for density distributions of
the neutrons and protons in a nucleus. Recently \cite{Fattoyev12,Erler13},
the correlations of the neutron-skin thickness in $^{208}$Pb nucleus
with several bulk properties of neutron stars have been examined for
the TOV-min and FSU type models. The energy density functional for the
TOV-min corresponds to  the Skyrme type effective force and that for FSU
is based on the extended RMF model.  The  correlation between neutron-skin
thickness in $^{208}$Pb nucleus and the neutron star radius $R_{1.4}$
for TOV-min is noticeably smaller than the one obtained for the FSU model.
Consequently, for the case of TOV-min the properties of neutron stars can
have larger variations at a fixed neutron-skin in $^{208}$Pb nucleus.
This result is in concordance with the large uncertainties in the high
density behaviour of the symmetry energy for the Skyrme type energy
density functionals \cite{Dutra12}.

The tighter correlations of neutron-skin thickness in $^{208}$Pb nucleus
with the several properties of the neutron stars within the RMF models
seem to be stemming from the lack of freedom in the isovector part of
the effective Lagrangian density associated with these models.  Most of
the RMF models describe the density dependence of the symmetry energy
either in terms of the  coupling of the isovector-vector $\rho$ mesons
with the nucleons \cite{Lalazissis97} or by including  only an additional
cross-coupling of  isosclar-scalar  $\sigma$ or  isosclar-vector $\omega$
with the isovector-vector  $\rho$ mesons \cite{Estal01,Todd-Rutel05}. The
$\sigma-\rho$ or $\omega-\rho$ cross-coupling allows one to  vary
the neutron-skin thickness over a wide range without significantly
affecting the quality of fit to the bulk properties of the nuclei such as
the  total binding energy and the charge radii\cite{Horowitz01}.  The  differences in the
high density behaviour of the symmetry energy and their consequences on
the properties of the neutron stars arising due to the use of different
cross-couplings have never been studied in detail. Further, the
inclusion of isovector-scalar $\delta$ mesons can modify the  behaviour
of the symmetry energy at high densities 
\cite{Liu02,Roca-Maza11,Klahn06}.  In principal,
the contributions from the  various cross-couplings and the $\delta$
mesons should be included in a single RMF model.  It has not been done so
far due to lack of accurate experimental data on the finite nuclei and
the  neutron stars which govern the isovector part of the RMF model.
A comprehensive study of variations in the properties of the neutron
stars as a function of the neutron-skin thickness  within  the RMF models
corresponding to different choices for the cross-coupling terms with and
without the inclusion of the $\delta$ mesons may be highly desirable.
Such investigation would enable one to understand to what extent the
inclusion of contributions from these various cross-couplings and the
$\delta$ mesons within a single RMF model are  necessary to describe
simultaneously the neutron-skin thickness in $^{208}$Pb and the various
neutron star properties.

In the present work, we would like to study the diversities in the
properties of the neutron stars arising purely due to the uncertainties
in the isovector part of the effective Lagrangian density which governs
the density dependence of the nuclear symmetry energy in the extended
RMF model.  Towards this purpose, two different families of extended RMF
models are obtained which mainly differ from each other in the choice for
the cross-coupling term in the isovector part of the effective Lagrangian
density.  One of the families of models includes $\sigma - \rho$
cross-coupling while the other includes $\omega - \rho$ cross-coupling
term in addition to the various linear and non-linear interaction terms
already present in the commonly used RMF models.  The contributions due
the coupling of  the $\delta$ mesons to the nucleons are also considered.
The various coupling constants are so varied that they produce wide
variations in the neutron-skin thickness in $^{208}$Pb nucleus without
affecting significantly the binding energies and charge radii of the
finite nuclei.  The various neutron star properties considered are
the core-crust transition density, radius for the neutron stars with canonical
mass, the tidal polarizability parameter and the threshold mass
required for the enhanced cooling through direct Urca process.  Some of
these neutron stars properties at a fixed neutron-skin thickness differ
significantly for two different families of the models.

The paper is organized as follows.  In Sec.  II, we briefly outline the
form of the effective Lagrangian density for the extended RMF model.
The procedure adopted to  obtain different parameterizations  for
two different families of the  model are described in Sec. III. We
also present in this section the results depicting  the relationship
between different coupling constants which govern  the isovector part
of the effective Lagrangian density. In Sec. IV, the various properties
of the neutron stars obtained for the different families of models are
compared for fixed values of neutron-skin thickness in $^{208}$Pb nucleus.
The main conclusions  are presented in Sec. V.

\section{ Theoretical frame work }
\label{sec:framework}

In the RMF models, nucleons interact through the exchange of isoscalar
scalar $\sigma$, isoscaler vector $\omega$ and isovector vector $\rho$
mesons.  The  effective Lagrangian density  for the RMF model usually
includes the cubic and quartic order non-linear self-interaction terms
for the $\sigma$ mesons in addition to the linear terms for the $\sigma$,
$\omega$ and $\rho$ mesons which describe their interactions with the
nucleons. The non-linear self-interaction terms for the $\sigma$
mesons are added to yield reasonable values for the empirical
properties of symmetric nuclear matter. Further, the RMF models
are extended by including various cross-couplings terms for these
mesons and self-interaction terms for the $\omega$ and $\rho$ mesons.
The $\sigma-\rho$ and $\omega-\rho$ cross-coupling  terms enables one
to vary the density dependence of the symmetry energy coefficient and
the neutron skin thickness in heavy nuclei over a wide range without
affecting the other properties of finite nuclei \cite{Furnstahl02,Sil05}.
Most of the RMF models do not include the contributions from the
isovector-scalar $\delta$ mesons.  The  bulk properties of the finite
nuclei like binding energies and radii are not very sensitive to the
presence of the $\delta$ mesons.  However, earlier investigations have
stressed  the need to include the contributions from the $\delta$ mesons
for proper description of the highly  asymmetric dense matter.
\cite {Huber96,Kubis97,Liu02,Greco03,Klahn06}.

The effective  Lagrangian density which  includes  the
lowest order contribution from the $\delta$ mesons together with the
various non-linear cross-coupling and self-interaction  contributions
already present in the extended RMF model, can be written as,
\cite{Furnstahl96,Serot97,Furnstahl97,Kubis97,Liu02},

 \begin{equation}
\label{eq:lden}
{\cal L}= {\cal L_{NM}}+{\cal L_{\sigma}} + {\cal L_{\omega}} + {\cal
L_{\mathbf{\rho}}} + {\cal L_{\delta}}+{\cal L_{\sigma\omega\mathbf{\rho}}}, 
\end{equation}
where the Lagrangian ${\cal L_{NM}}$ describing the interactions of the
nucleons through the mesons is, 
\begin{eqnarray}
\label{eq:lbm}
{\cal L_{NM}} &=& \sum_{J=n,p} \overline{\Psi}_{J}[i\gamma^{\mu}\partial_{\mu}- (M-g_{\sigma} \sigma- g_{\delta}\delta \tau)\nonumber\\
&&-(g_{\omega }\gamma^{\mu} \omega_{\mu}+\frac{1}{2}g_{\mathbf{\rho}}\gamma^{\mu}\tau .\mathbf{\rho}_{\mu})]\Psi_{J}. 
\end{eqnarray}
Here, the sum is taken over the neutrons and protons and 
$\tau$ are the isospin matrices. The Lagrangian density ${\cal L}_i$
for $i= \sigma, \omega, \rho$   and $\delta$  can be written as,
\begin{equation}
\label{eq:lsig}
{\cal L_{\sigma}} =
\frac{1}{2}(\partial_{\mu}\sigma\partial^{\mu}\sigma-m_{\sigma}^2\sigma^2)
-\frac{{\kappa_3}}{6M}
g_{\sigma}m_{\sigma}^2\sigma^3-\frac{{\kappa_4}}{24M^2}g_{\sigma}^2 m_{\sigma}^2\sigma^4,
\end{equation}
\begin{equation}
\label{eq:lome}
{\cal L_{\omega}} =
-\frac{1}{4}\omega_{\mu\nu}\omega^{\mu\nu}+\frac{1}{2}m_{\omega}^2\omega_{\mu}\omega^{\mu}+\frac{1}{24}\zeta_0 g_{\omega}^{2}(\omega_{\mu}\omega^{\mu})^{2},
\end{equation}
\begin{equation}
\label{eq:lrho}
{\cal L_{\mathbf{\rho}}} =
-\frac{1}{4}\mathbf{\rho}_{\mu\nu}\mathbf{\rho}^{\mu\nu}+\frac{1}{2}m_{\rho}^2\mathbf{\rho}_{\mu}\mathbf{\rho}^{\mu},
\end{equation}
\begin{equation}
\label{eq:ldel}
{\cal L_{\delta}} =
\frac{1}{2}(\partial_{\mu}\delta\partial^{\mu}\delta-m_{\delta}^2\delta^2).
\end{equation}
The $\omega^{\mu\nu}$, $\mathbf{\rho}^{\mu\nu}$ are field tensors
corresponding to the $\omega$ and $\rho$ mesons, and can be defined as
$\omega^{\mu\nu}=\partial^{\mu}\omega^{\nu}-\partial^{\nu}\omega^{\mu}$
and $\mathbf{\rho}^{\mu\nu}=\partial^{\mu}\mathbf{\rho}^{\nu}-
\partial^{\nu}\mathbf{\rho}^{\mu}$.  The cross interactions of
$\sigma, \omega$, and $\mathbf{\rho}$ mesons are described by ${\cal
L_{\sigma\omega\rho}}$ which can be written as,

 \begin{equation}
\label{eq:lnon-lin}
\begin{split}
{\cal L_{\sigma\omega\rho}} & =
\frac{\eta_1}{2M}g_{\sigma}m_{\omega}^2\sigma\omega_{\mu}\omega^{\mu}+ 
\frac{\eta_2}{4M^2}g_{\sigma}^2 m_{\omega}^2\sigma^2\omega_{\mu}\omega^{\mu}\\
&+\frac{\eta_{\rho}}{2M}g_{\sigma}m_{\rho }^{2}\sigma\rho_{\mu}\rho^{\mu} 
+\frac{\eta_{1\rho}}{4M^2}g_{\sigma}^2m_{\rho }^{2}\sigma^2\rho_{\mu}\rho^{\mu}\\
&+\frac{\eta_{2\rho}}{4M^2}g_{\omega}^2m_{\rho}^{2}\omega_{\mu}\omega^{\mu}\rho_{\mu}\rho^{\mu}.
\end{split}
\end{equation}
One also needs to include the contributions from the electromagnetic
interaction in the case of finite nuclei. The Lagrangian density ${\cal
L}_{em}$ for the electromagnetic interaction  can be written as,
\begin{equation}
\label{eq:lem}
{\cal L}_{em}= -\frac{1}{4}F_{\mu\nu}F^{\mu\nu}- 
e\overline{\Psi} _{p}\gamma_{\mu}A_{\mu}\Psi_{p},
\end{equation}
where, $A$ is the photon filed and
$F^{\mu\nu}=\partial^{\mu}A^{\nu}-\partial^{\nu}A^{\mu}$.  The equation
of motion for nucleons, mesons and photons can be derived from the
Lagrangian density defined in Eq.(\ref{eq:lden}).  The contributions
from Eq. (\ref{eq:lem}) are included only for the case of finite nuclei.

It is clear from (\ref{eq:lnon-lin}) that there are five cross-coupling
terms; two of them are the cubic order terms corresponding to the
$\sigma-\omega$ and $\sigma-\rho$ cross-couplings and the remaining are
the  quartic order terms.  The contribution from the $\sigma-\omega$
cross-couplings and self coupling of $\omega$ mesons play important
role in varying the high density behaviour of the EOSs and also
prevents instabilities in them \cite{Sugahara94,Estal01,Muller96}.
The contributions of the self-coupling of $\rho$ mesons are not considered
in Eq. (\ref{eq:lrho}),  since, expectation value of the $\rho$ meson
field is order of magnitude smaller than that for the $\omega$ meson field
\cite{Serot97}. The inclusion of the $\rho$ meson self interaction can
affect the properties of the finite nuclei and neutron stars only very
marginally \cite{Muller96}.  Of the particular interest in the present
work are the cross-coupling terms involving $\rho$ meson field which
contributes to the isovector part of the effective Lagrangian density in
addition to the usual linear couplings of the $\rho$ and $\delta$ mesons
to the nucleons.  We shall mainly focus on the lowest order $\sigma-\rho$
and $\omega-\rho$  cross-couplings  whose strengths  are determined by the
values of $\eta_{\rho}$ and $\eta_{2\rho}$.  The quartic order $\sigma-\rho$
cross-coupling strength $\eta_{1\rho}$ is set to zero.   The values of
$\eta_\rho$ or $\eta_{2\rho}$ can be appropriately adjusted to yield  wide
variations in the density dependence of the symmetry energy coefficient
and the neutron skin thickness in heavy nuclei  without affecting the
other properties of finite nuclei \cite{Furnstahl02,Sil05,Dhiman07}.

\section{Model parameters} 

Two different families of the extended RMF models, named hereafter as
$F_\rho$ and $F_{2\rho}$, are obtained. These families differ from each other 
in the choice for the cross-coupling term in the isovector part of the
Lagrangian density. The  isovector part of the Lagrangian density for
the $F_\rho$($F_{2\rho}$) family is governed by the coupling parameters $g_\rho$,
$g_\delta$ and $\eta_\rho$($\eta_{2\rho}$). The parameters $g_\rho$ and $g_\delta$
denote the strengths for the coupling of the $\rho$ and $\delta$ mesons
to the nucleons, respectively. The parameter $\eta_\rho$ and $\eta_{2\rho}$ denote the
strength of the $\sigma - \rho$ and $\omega - \rho$ cross-couplings as can be seen from Eq.
\ref{eq:lnon-lin}. The remaining parameters which correspond to the
isoscalar part of the Lagrangian density and the mass of the $\sigma$,
$\omega$ and $\rho$ mesons are kept fixed to that of the BKA22 model
\cite{agrawal10}.  The BKA22 model has been identified to satisfy various
constraints related to symmetric nuclear matter, pure neutron matter,
symmetry energy, and its derivatives \cite{Dutra14}.

\begin{table}
\caption{\label{tab1}
The parameters of the isovector part of the Lagrangian density  for some
representative sets for the $F_\rho$ and $F_{2\rho}$ families of models.
In the bottom part, the values for  the symmetry energy coefficient
at the saturation density $J$, symmetry energy slope parameter $L$,
effective mass for the protons and neutrons and their differences are
also presented.
 All these quantities  are in MeV.
} \begin{tabular}{|c|cccc|} \hline \multicolumn{1}{|c}{}&
\multicolumn{2}{|c}{$F_\rho$}& \multicolumn{2}{c|}{$F_{2\rho}$}\\
\cline{2-5} Parameter& SET1&  SET2& SET3&SET4\\ \hline
 $\eta_\rho$         &        4.0  &   4.0  & 0.0  &   0.0\\
 $\eta_{2\rho}$      &         0.0   &   0.0 &  17.5&   17.5\\ $g_\delta$
 &     0.0  &     8.0 &  0.0   &   8.0\\ $g_\rho$            &    13.033&
 21.863&  11.051&  18.556\\
\hline
 $J$               &   33.3&  30.9 &    33.0&    30.2\\ $L$
 &  79.0 &   62.5&     65.0&   23.3  \\
$M^*_p$            &    570.1  &  630.9      & 577.8&   652.4\\ $M^*_n$
&    570.1 & 495.4         &577.8  &     515.8\\ $\Delta M^*_{pn}$  &
0.0      &   135.5 &  0.0&   136.6\\
 \hline
\end{tabular} \end{table}
 \begin{table}
\caption{\label{tab2} The values of the total binding energy ($E$ )
in MeV, charge radii ($r_c$), neutron radii ($r_n$) and neutron-skin
thickness $\Delta r_{\rm np}$ in fm for a few asymmetric spherical nuclei
obtained for SET1 - SET4 parameters.}

 \begin{tabular}{|cc|cccc|}
 \hline
\multicolumn{2}{|c}{}&
\multicolumn{2}{|c}{$F_\rho$}&
\multicolumn{2}{c|}{$F_{2\rho}$}\\
\cline{3-6}
 Nucleus&Property&   SET1& SET2& SET3& SET4\\
\hline
$^{48}{\rm Ca}$&$E$&   -415.75 &   -415.40&  -415.81&  -415.61\\  
&$r_c$&               3.468 &   3.484& 3.465& 3.477 \\
&$r_n$&               3.575  &  3.544 & 3.574& 3.535 \\
&$\Delta r_{\rm np}$&          0.201 &   0.153&  0.202& 0.151 \\
\hline
$^{132}{\rm Sn}$&$E$&   -1102.49 &  -1100.37 &   -1102.69&     -1100.99 \\
&$r_c$&                 4.736 &  4.759 &  4.727&   4.731   \\
&$r_n$&                   4.952  &4.901  &4.934 & 4.873 \\
&$\Delta r_{\rm np}$&         0.284 &  0.210 & 0.286& 0.210  \\
\hline 
$^{208}{\rm Pb}$&$E$&  -1637.07  & -1637.08&  -1637.06 & -1637.05  \\
&$r_c$&                      5.545&  5.566 & 5.535  &5.538\\
&$r_n$&                        5.706 & 5.659  &5.699  &5.631\\
&$\Delta r_{\rm np}$&                 0.219 &   0.151  & 0.221   &  0.151 \\
\hline
\end{tabular}
\end{table}

The different parameterizations  of $F_\rho$($F_{2\rho}$) families are
obtained by  varying appropriately the  values of $g_{\rho}$, $g_\delta$
and $\eta_\rho(\eta_{2\rho})$. For a given value of $g_\delta$ and
$\eta_\rho(\eta_{2\rho})$,   the value of $g_\rho$ is always adjusted to
yield appropriate binding energy for the $^{208}$Pb nucleus.  Once the
values of $g_\delta$, $g_\rho$ and $\eta_\rho$ or $\eta_{2\rho}$ are
known, the  properties of the nuclear matter and the finite nuclei can
be computed.  We vary the values of $g_\delta$ over a wide range from 0
to 8. The values of $\eta_\rho$ and $\eta_{2\rho}$ are varied in the
range of $0 - 12$ and $0 - 60$, respectively. For $\eta_\rho > 12$,
the stable solutions of the field equations for the mesons could not
be obtained. We have constructed 22 different parameterizations of the
$F_\rho$ and 41 different parameterizations of the $F_{2\rho}$ families.
\begin{figure}
\includegraphics[width=0.95\columnwidth,angle=0,clip=true]{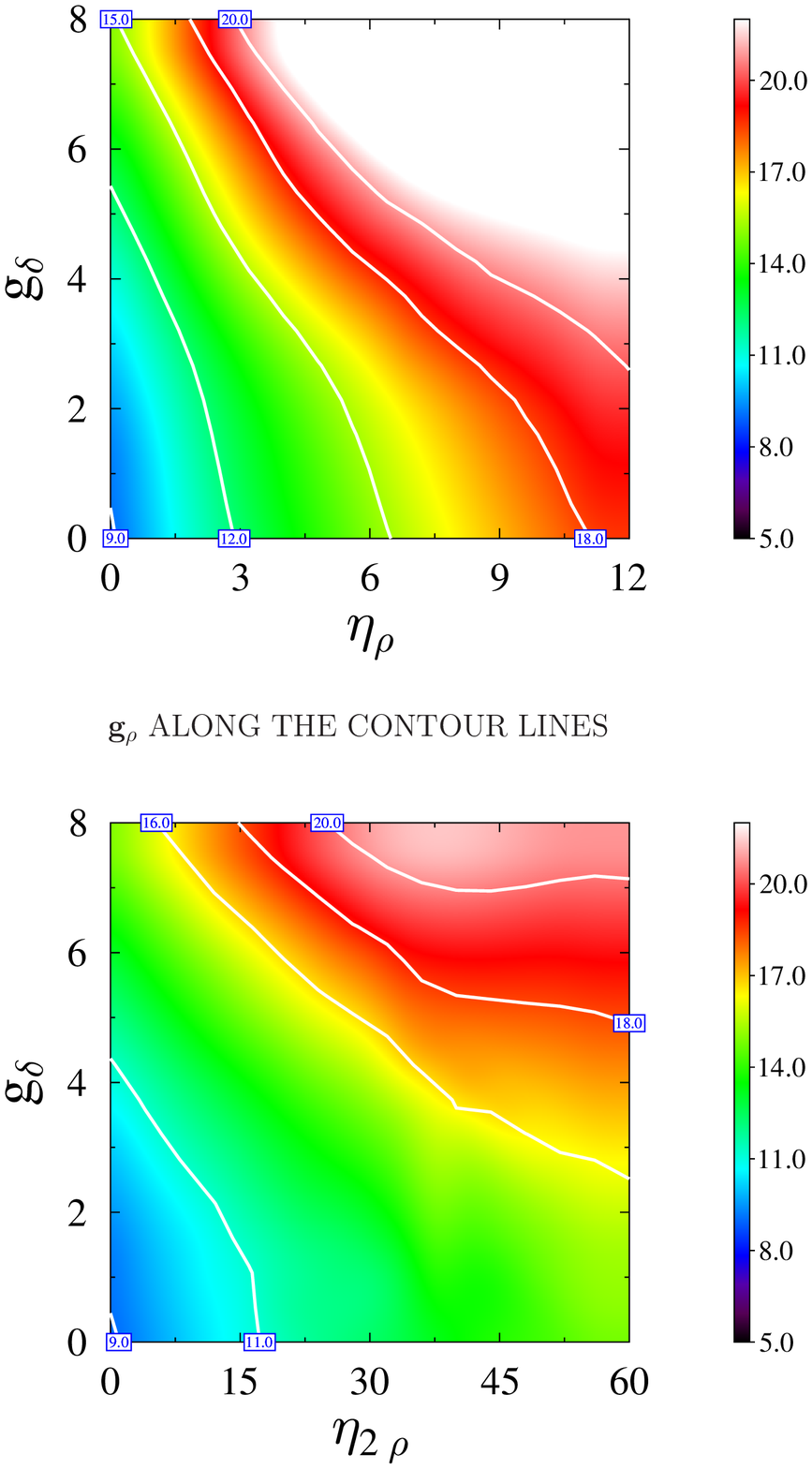}
\caption{(Color online) Colour coded contours in the $g_\delta -
\eta_{\rho}$(upper panel)  and $g_\delta - \eta_{2\rho}$(lower panel) planes 
corresponding to the $F_{\rho}$ and $F_{2\rho}$ families, respectively.  The value of $g_\rho$ are colour coded according
to the scale one the right side.  {\label{fig1} }}
\end{figure}

\begin{figure}
\includegraphics[width=0.95\columnwidth,angle=0,clip=true]{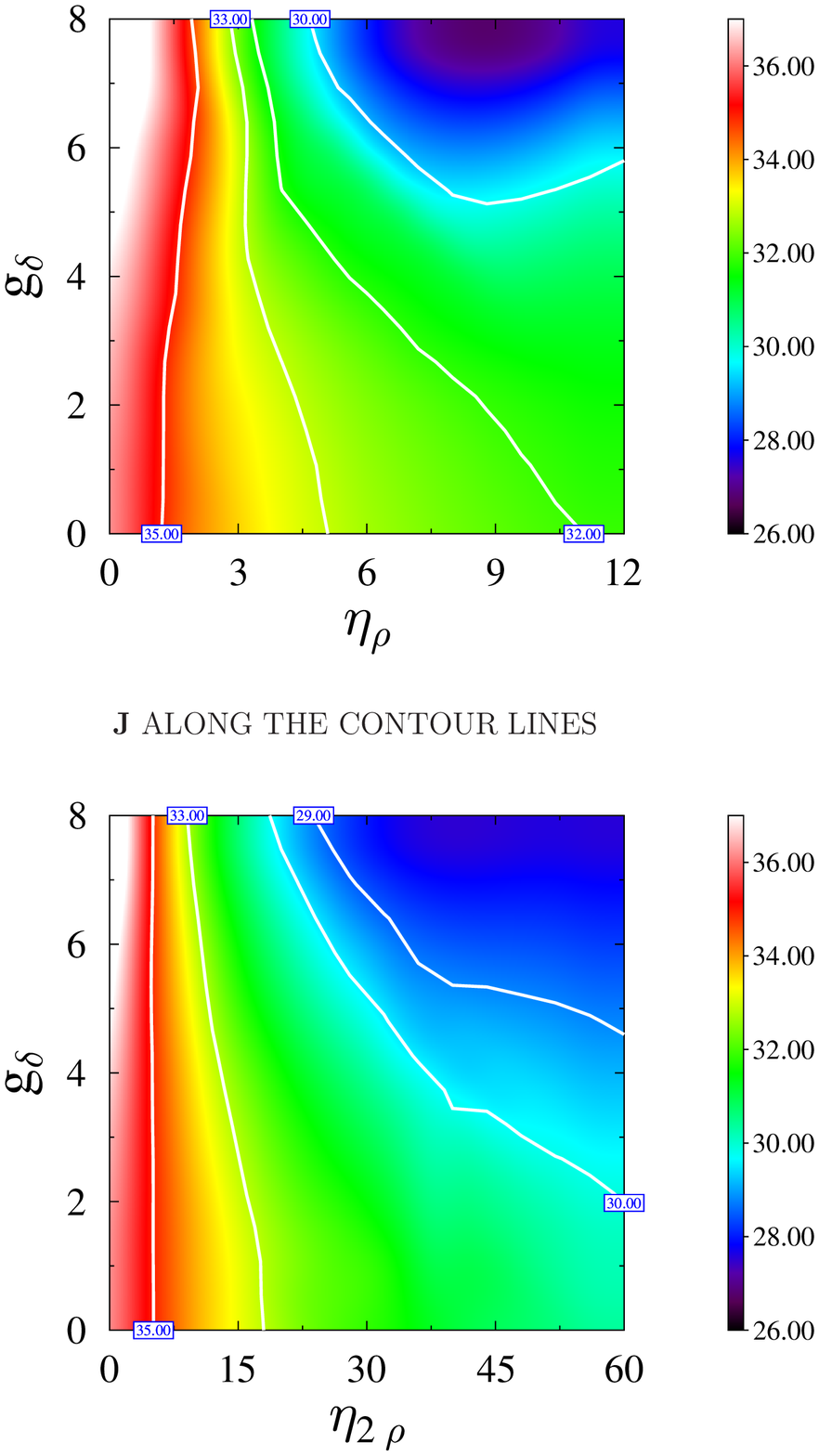}
\caption{(Color online)Same as Fig. \ref{fig1}, but, for the symmetry
energy at the saturation density $(J = S(\rho_0))$ fixed along
the contour.  The values of $J$ are in MeV. {\label{fig2} }}
 \end{figure}

\begin{figure}
\includegraphics[width=0.95\columnwidth,angle=0,clip=true]{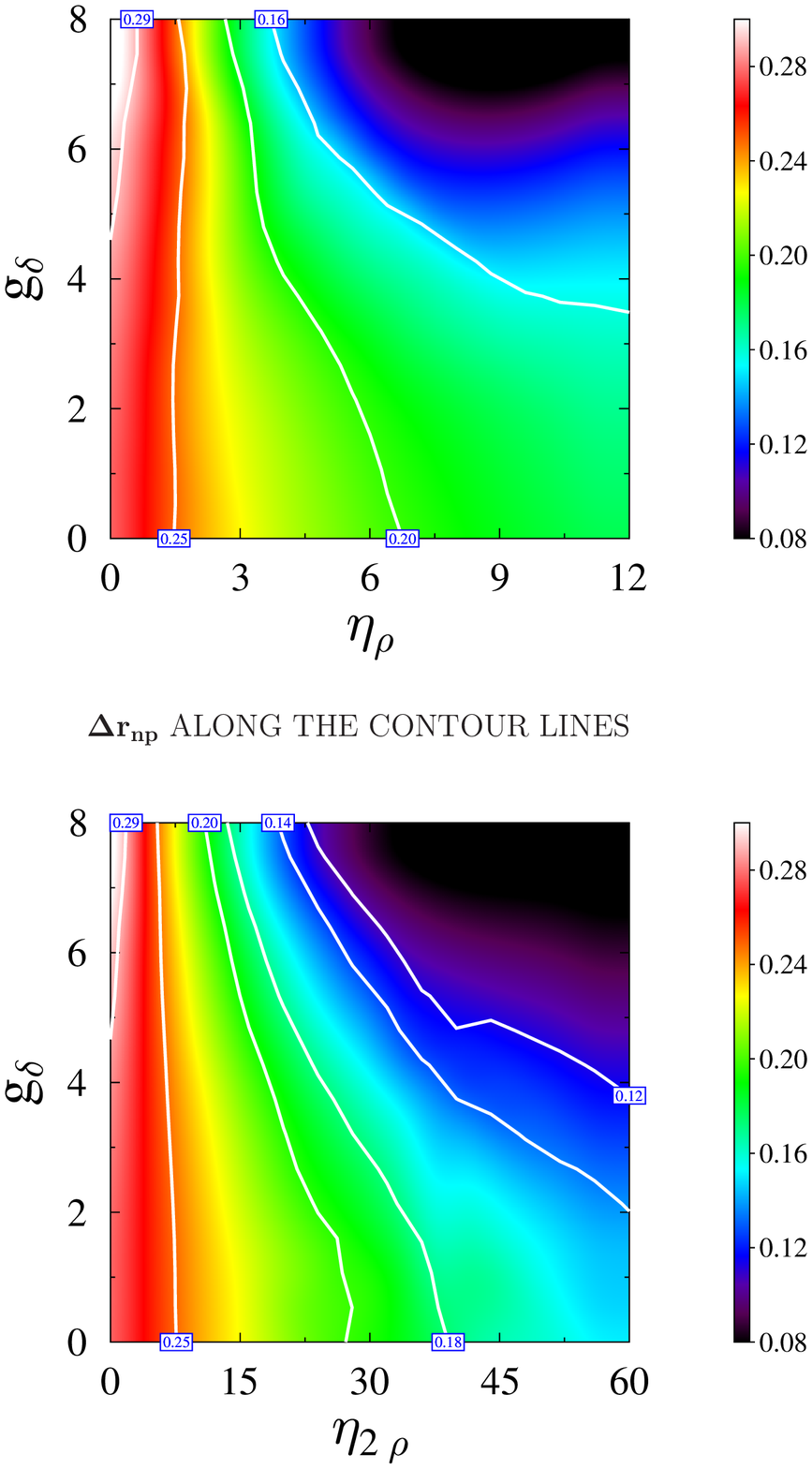}
\caption{(Color online)Same as Fig. \ref{fig1}, but, for the  neutron-skin
thickness $\Delta r_{\rm np}$ in the $^{208}$Pb nucleus fixed along
the contour.  The values of $\Delta r_{\rm np}$ are in  fm. {\label{fig3} }}
 \end{figure}

The  Fig. \ref{fig1} displays the relationship between the various
parameters of the isovector channel for the $F_\rho$ and $F_{2\rho}$
families of the models.  It can be readily seen that the value of
$g_\rho$, required to reproduce the binding energy for the $^{208}$Pb
nucleus, increases with $g_\delta$, $\eta_\rho$ and $\eta_{2\rho}$. In
other words,the equation of state at least for the densities relevant for
the finite nuclei becomes softer with the increase in the $g_\delta$,
$\eta_\rho$ and $\eta_{2\rho}$, which is compensated by increasing the
value of $g_\rho$ to reproduce the binding energy for the $^{208}$Pb
nucleus.  In the Fig. \ref{fig2}, the relationship of the parameters
$g_{\delta}$ and $\eta_{\rho}$($\eta_{2\rho}$) with the symmetry
energy coefficient at the saturation density ($J=S(\rho_0)$) for the
$F_\rho$($F_2\rho$) is displayed in terms of the contour plots. Similarly,
the results for the $\Delta r_{\rm np}$ in the $^{208}$Pb nucleus
are plotted in Fig.  \ref{fig3}.  In general, the values of $J$ and
$\Delta r_{\rm np}$ decreases with increasing $g_\delta$, $\eta_\rho$
or $\eta_{2\rho}$.  The $F_{2\rho}$ model yields  larger variations in
$\Delta r_{\rm np}$. The large values for $\eta_\rho$ are not favored,
as a result the $F_\rho$ family can yield very small values of
$\Delta r_{\rm np}$ only with the inclusion of the $\delta$ mesons.
In Table \ref{tab1} the values of the parameters for four representative sets  corresponding
to the $F_\rho$ and $F_{2\rho}$ models are listed. 
The SET1 and SET2 belong to the $F_\rho$ family, while, SET3 and SET4
are for the $F_{2\rho}$ family.  The SET1 and SET3 do not  include the
contributions from the $\delta$ mesons ($g_\delta = 0$). The SET2 and
SET4 correspond to the highest value of the $\delta$-nucleon coupling
strength ($g_\delta = 8$), otherwise, they are very much similar to
the SET1 and SET3, respectively.  These different sets are so chosen
that the comparison of the properties of the neutron stars resulting
from them would give  us a crude estimate about the effects of $\delta$
meson as well as the $\sigma-\rho$ and $\omega-\rho$ cross-couplings.
In the bottom part of Table  \ref{tab1}, the values of the symmetry
energy coefficient at the saturation density ($J=S(\rho_0)$), symmetry
energy slope parameter ($L=\left (3\rho\frac{\partial S(\rho)}{\partial
\rho}\right )_{\rho_0}$), proton and neutron effective masses and their
differences are also presented. The effective masses are obtained at the
maximum asymmetry, i.e., the  pure neutron matter.  In Table \ref{tab2},
we present some bulk properties of a few asymmetric spherical nuclei. The
various bulk properties for these nuclei are relatively better reproduced
for the SET1 and SET3 parameters which corresponds to $\Delta r_{\rm
np} \sim 0.22$fm in the $^{208}$Pb nucleus, since, this value of $\Delta
r_{\rm np}$ is almost the same as that of the base model BKA22.  It may be
noted that the $\Delta r_{\rm np} = 0.22$ fm for the SET1 and SET3, but,
they belong to different families.  Similarly, SET2 and SET4 represent
different families, with  $\Delta r_{\rm np} = 0.15$ fm.


\begin{figure}
\includegraphics[width=0.8\columnwidth,angle=0,clip=true]{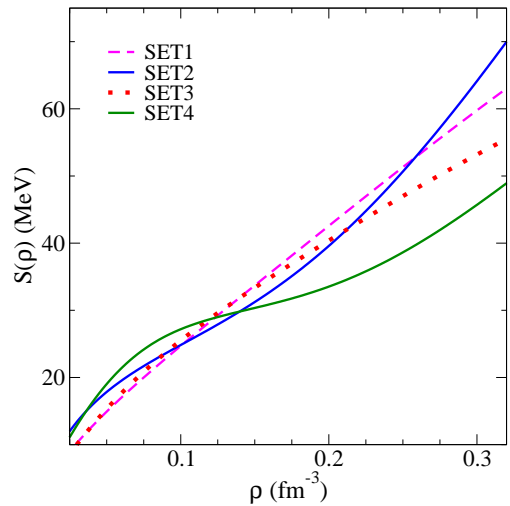}
\caption{(Color online)
The density dependence of symmetry energy $S(\rho)$ for
some representative cases of $F_\rho$ and $F_{2\rho}$ families of
the models. The labels SET1 and SET2 correspond to the  two different
parameterizations for the $F_\rho$ family, whereas, the SET3 and SET4
correspond to the $F_{2\rho}$ family. The SET1 and SET3 are associated
with $\Delta r_{\rm np} = 0.22 \text{fm}$ and the SET2 and SET4 yield
$\Delta r_{\rm np} = 0.15 \text{fm}$ (see also Tables \ref{tab1} and \ref{tab2}).}
{\label{fig4} }
\end{figure}

Let us now take a look at the density dependence of the symmetry energy
for the different parameterizations of the $F_\rho$ and
$F_{2\rho}$ families corresponding to equal values of $\Delta r_{\rm np}$.
In Fig. \ref{fig4}, we display the variations of symmetry energy as
a function of density for different parameterizations as indicated by
SET1, SET2, SET3 and SET4.  The high density behaviour for the symmetry
energy is stiffer for the $F_\rho$ family as can be easily verified by
comparing the results for the SET1 and SET2 with those for the SET3 and
SET4, respectively.  Further, by comparing the results for the SET1
with SET2 or those for SET3 with SET4, it can be concluded that the
inclusion of the $\delta$ mesons softens the symmetry energy at low
densities  while makes it  stiffer at higher densities.  The results
depicted in Fig. \ref{fig4} provide evidences a priori  about the
possibilities of the differences in the properties of neutron stars at a
fixed $\Delta r_{\rm np}$ across the different families of the models, due
to the differences in the high density behaviour of the symmetry energy.

\section{Neutron-skin thickness and properties of neutron stars}

We wish to study the differences in the properties of neutron stars 
for the $F_{\rho}$ and $F_{2\rho}$ families of the models at  fixed
values for the neutron skin thickness $\Delta r_{\rm np}$ in the
$^{208}$Pb nucleus.  In particular,  attention is given to the study of
such differences
 at $\Delta r_{\rm np} = 0.15$fm in $^{208}$Pb nucleus. This value of
 $\Delta r_{\rm np}$
is consistent with $0.156^{+0.025}_{-0.021}$ fm \cite{Tamii11}
and $0.168\pm 0.022$ fm \cite{Piekarewicz12} as extracted from the
experimental data on the dipole polarizability for $^{208}$Pb nucleus.  A
very recent measurement of coherent pion photo-production \cite{Tarbert14}
also corresponds to $\Delta r_{\rm np} = 0.15 \pm 0.03$ fm  in $^{208}$Pb
nucleus.  However, these measurements do not conclusively yet rule out the
larger values for $\Delta r_{\rm np}$, since, the Lead Radius Experiment
(PREX) \cite{Horowitz01,Abrahamya12,Donnelly89} has recently measured
$\Delta r_{\rm np}= 0.33^{+0.16} _{-0.18}$ fm  in $^{208}$Pb nucleus
via parity-violating electron scattering which provides the first purely
electroweak, almost model independent estimate.  Our purpose is to give
quantitative estimates about the extent to which the   various properties
of the neutron stars might vary across the different families of the
models  for a plausible value of  neutron-skin thickness.  The various
properties for the neutron stars considered are the core-crust transition
density, radius, red-shift, the threshold mass required for the enhance
cooling through the direct Urca process and the tidal polarizability
parameter. The comparison of results for the $F_{\rho}$ and $F_{2\rho}$
families of the models would enable us to understand the role of different
cross-coupling terms. We shall also assess the  effects of the $\delta$
mesons by comparing the results obtained  with and without its inclusion
in the same family of the models.

\begin{figure}
\includegraphics[width=0.8\columnwidth,angle=0,clip=true]{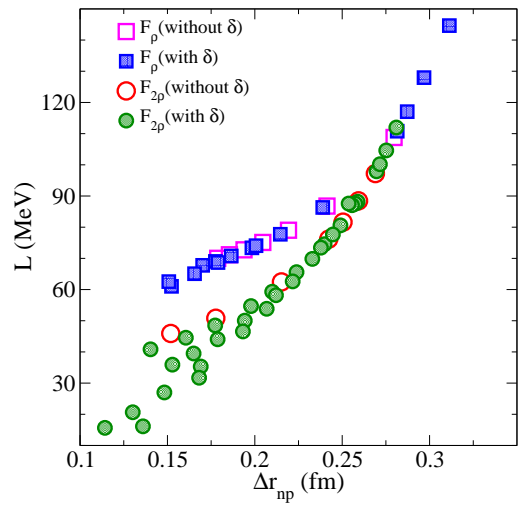}
\caption{(Color online)
The variations of symmetry energy slope parameter $L$ with neutron-skin
thickness $\Delta r_{\rm np}$ in the $^{208}$Pb nucleus for the $F_\rho$
and $F_{2\rho}$ families of models.  The solid and hollow symbols
represent the results obtained with ($g_\delta \not=0)$ and with
($g_\delta =0)$, respectively.  {\label{fig5} }}
\end{figure}

The values of the neutron-skin thickness in a heavy nucleus, according
to the Droplet Model \cite{Myers80}, are strongly correlated with the
symmetry energy slope parameter $L$. The dependence of $L$ on $\Delta
r_{\rm np}$ in $^{208}$Pb for the $F_{\rho}$ and $F_{2\rho}$ families of
the models are displayed in Fig.  \ref{fig5}.
It may be pointed out that the similar values of $\Delta r_{\rm np}$
can be obtained within a  given family by varying appropriately the values of coupling
parameters $g_\delta$ and $\eta_\rho$ or $\eta_{2\rho}$(see also Fig. \ref{fig3}).
  The solid and the hollow
symbols represent the results obtained with and
without  the contributions from the $\delta$ mesons, respectively.
The  values of $\Delta r_{\rm np}$ are well
correlated with $L$ within a given family of the models irrespective
of the contributions from the $\delta$ mesons.  However, the values of
$L$ for the two families of the models differ significantly at smaller
$\Delta r_{\rm np}$. This difference  gradually disappears as $\Delta
r_{\rm np}$ increases.

\begin{figure}
\includegraphics[width=0.8\columnwidth,angle=0,clip=true]{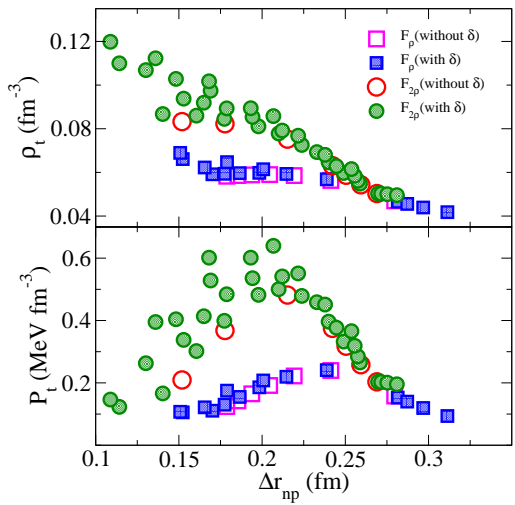}
\caption{(Color online)
The variation of core-crust transition density and the corresponding
pressure  with the neutron-skin thickness $\Delta r_{\rm np}$ in
$^{208}$Pb nucleus for the $F_\rho$ and $F_{2\rho}$  families of the
extended RMF models.
{\label{fig6} }}
\end{figure}

The values of core-crust transition density $\rho_t$ and the
corresponding pressure $P_t$ as a function of $\Delta r_{\rm np}$
obtained for $F_\rho$ and $F_{2\rho}$ families of models are
plotted in Fig. \ref{fig6}.  The values of $\rho_t$ are obtained
using a method based on the relativistic random-phase approximation
\cite{Carriere03,Sulaksono09,Sulaksono07,Piekarewicz07}.  This method
uses the fact that the uniform matter in  its ground state at sufficiently
low densities becomes unstable to small density fluctuations.  The values
of $\rho_t$ are correlated with the $\Delta r_{\rm np}$ within a same
family irrespective of the contributions from the $\delta$ mesons. But,
this correlation seems to be some what model dependent--- the values
of $\rho_t$ for both the families of models at a fixed $\Delta r_{\rm
np}$ are not the same.  In particular, the $\rho_t$ is  significantly
larger for the $F_{2\rho}$ family at smaller $\Delta r_{\rm np}$.  
The transition pressure $P_t$ is not very well correlated with the $\Delta
r_{\rm np}$. Initially, the $P_t$ increases with $\Delta r_{\rm np}$
and it decreases for higher values of  $\Delta r_{\rm np}$.

\begin{figure}
\includegraphics[width=0.8\columnwidth,angle=0,clip=true]{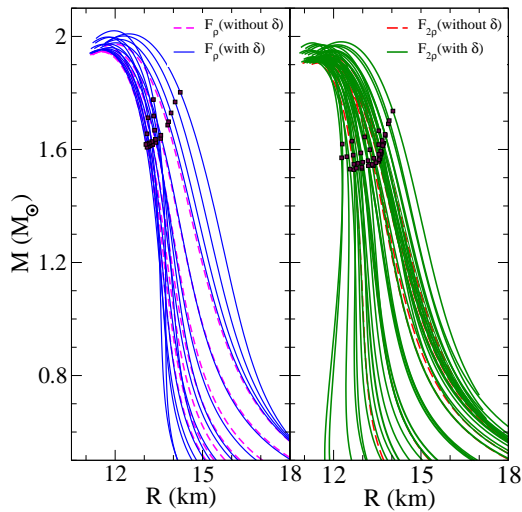}
\caption{(Color online)
The mass-radius relationship for the $F_\rho$ and $F_{2\rho}$ families
of the models.  The solid squares represent the masses and the corresponding radii for the neutron stars
with the central density to be $3\rho_0$.  {\label{fig7} }}
 \end{figure}

Once, the core-crust transition density is determined, the EOS for various
density ranges as required for the computation of the properties of the
neutron stars can be constructed. The  EOS  data for the density $\rho
\sim 4.8\times 10^{-9} - 2.6\times 10^{-4} \text{fm}^{-3}$ corresponding
to the outer crust region are taken from Ref. \cite{Ruster06}.
The EOS for the inner crust is obtained by assuming a polytropic form
$P(\epsilon)=a+b\epsilon^{4/3}$, where $P$ and $\epsilon$ are the
pressure and energy density respectively.  The constants $a$ and $b$
are determined in such a way that the EOS for the inner crust  matches
with that for the inner edge of the outer crust at one end and with the
edge of the core at the other end.  The EOSs for the  core region, $\rho
> \rho_t$, are obtained within the RMF model by using the different
parameterizations of the $F_\rho$ and $F_{2\rho}$ families. The core
region is assumed to be composed of neutrons, protons, electrons and
muons. The chemical potentials for various particle species at a given
baryon density are obtained by imposing the $\beta$-equilibrium and charge
neutrality conditions. We use these EOSs to compute the  properties of
static  neutron stars by integrating the  Tolman-Oppenheimer-Volkoff
(TOV) equations \cite{Weinberg72}.  In Fig. \ref{fig7} we display
the mass-radius relationship for the sequences of static neutron stars
obtained for the $F_\rho$ and $F_{2\rho}$ families of models.
The solid and the dashed lines depict the results obtained with and 
without the inclusion of the contributions from the $\delta$ mesons, respectively.
The solid squares represent the masses and the corresponding radii for the neutron stars with the
central density to be $3\rho_0$.
The different EOSs obtained for a given family of the models differ 
mainly in the high density behaviour of the symmetry energy. This leads
to the variations in the mass-radius relationship for the neutron stars
within the same family of the models.The maximum mass  $M_{\rm
max} = 1.95 - 2.02 M_\odot$ and $1.91 - 1.98M_\odot$ and the radii
 $R_{1.4} = 13.3 - 15.4$km and $12.3- 14.9$km for the  $F_\rho$ and
 $F_{2\rho}$ families, respectively.
The value of $M_{\rm max}$ for the $F_\rho$ family is consistent with the
recent mass measurements $M = 1.97\pm 0.04M_\odot$ for PSR J1614-2230
\cite{Demorest10} and $M = 2.01\pm 0.04M_\odot$ for  PSR J0348+0432
\cite{Antoniadis13}, but, the values for $R_{1.4}$ is marginally away from
the $R_{1.4} = 10.7 - 13.1$km as extracted in Ref. \cite{Lattimer13}. For
the $F_{2\rho}$ family,  values for $M_{\rm max}$ are barely consistent
with the recent measurements, but, $R_{1.4}$ is consistent with the
ones extracted in Ref. \cite{Lattimer13}.  The inclusion of the $\delta$
mesons yields higher values for the maximum mass for the neutron stars
within a family.  A more realistic estimation for the effects of $\delta$
mesons on the maximum mass of the neutron stars requires the inclusion
of various exotic degrees of freedom. Since, the density at the center
of the neutron star with the maximum mass for our EOSs is significantly
larger than $3\rho_0$.

\begin{figure}
\includegraphics[width=0.8\columnwidth,angle=0,clip=true]{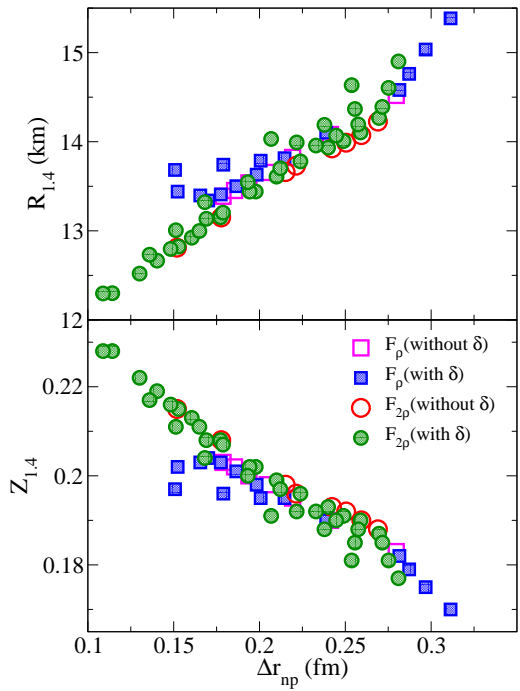}
\caption{(Color online)
Plots for the radius $R_{1.4}$  for the neutron stars with the canonical
mass $1.4M_\odot$ (upper panel) and the corresponding red shift (lower
panel)  as a function of $\Delta r_{\rm np}$ in the $^{208}$Pb nucleus
obtained for the $F_\rho$ and $F_{2\rho}$ families of models.
{\label{fig8} }} \end{figure}

\begin{figure}
\includegraphics[width=0.8\columnwidth,angle=0,clip=true]{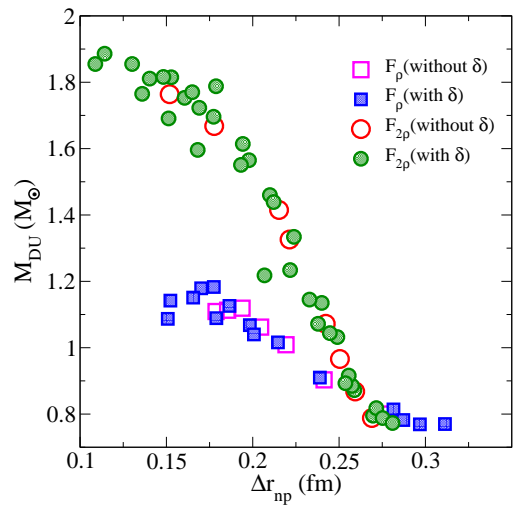}
\caption{(Color online)
The dependence of the threshold neutron star mass
$M_{\scriptscriptstyle {DU}}$ on the neutron-skin thickness in
$^{208}$Pb nucleus. The neutron stars with mass equal to or larger
than $M_{\scriptscriptstyle {DU}}$ undergo enhanced cooling through
direct Urca process for the cooling.  {\label{fig10} }} 
 \end{figure}

We now compare  the various properties of the neutron stars at fixed
values of $\Delta r_{\rm np}$ obtained for the $F_\rho$ and $F_{2\rho}$
families.  
Before embarking on our discussion, it may  be reminded  that the
dependence of the various neutron star properties on the neutron-skin
thickness are merely due to the fact that different models differ only in
the density dependence of the symmetry energy. The EOS for the symmetric
nuclear matter is taken to be  the same for all the models, since,
our goal is  to study the diversities in the properties of the neutron
stars arising purely due to the differences in the density dependence of
the symmetry energy within the extended RMF model.  In Fig. \ref{fig8}
the radii and red shifts for the neutron stars with mass  $1.4M_\odot$
are plotted against the $\Delta r_{\rm np}$.  The spread in $R_{1.4}$
and $Z_{1.4}$ for several cases corresponding to the similar $\Delta
r_{\rm np}$ within the same family is smaller.  The values of $R_{1.4}$
and $Z_{1.4}$ obtained for two different families differ noticeably at
the smaller values of $\Delta r_{\rm np}$. For $\Delta r_{\rm np} =0.15
$ fm the maximum  differences in the values of $R_{1.4}$ and $Z_{1.4}$
obtained for the two families are $\sim$ 1.0 km and 0.02, respectively.
In Fig. \ref{fig10} We display our results for threshold mass the
$M_{\scriptscriptstyle{DU}}$ required for the enhanced cooling of neutron
stars by means of neutrino emission from the nucleons in the direct Urca
process \cite{Lattimer91}.  The values of $M_{\scriptscriptstyle{DU}}$
are quite sensitive to the neutron-skin thickness. The value of
$M_{\scriptscriptstyle{DU}}$ for the $F_{2\rho}$ family  can  vary  over
the range of $0.8  - 1.9M_\odot$ with $\Delta r_{\rm np}$  decreasing
from 0.3fm to 0.1fm.   This variation is little smaller for the case of
$F_\rho$ family. The value of $M_{\scriptscriptstyle{DU}}$ for both the
families differ quite significantly at smaller  $\Delta r_{\rm np}$.
At  $\Delta r_{\rm np}$ =0.15 fm the  difference between the values of
$M_{\scriptscriptstyle{DU}}$ for both the families is about $0.6M_\odot$
which is quite significant ($40\%$).

\begin{figure}
\includegraphics[width=0.8\columnwidth,angle=0,clip=true]{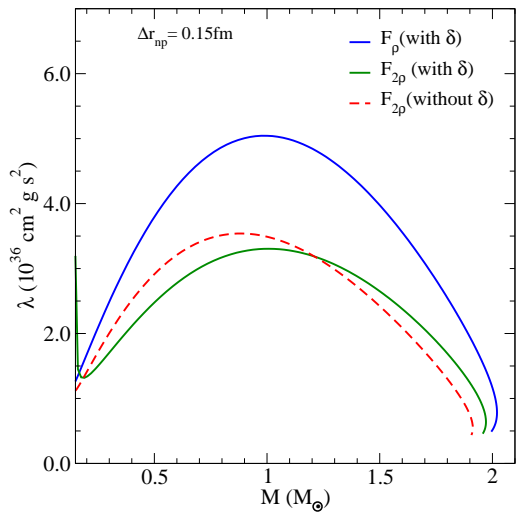}
\caption{(Color online)
Variations  in the tidal polarizability parameter $\lambda$  with the neutron star
mass for the  different parameterizations of $F_\rho$ and $F_{2\rho}$
families  corresponding to neutron-skin thickness $\Delta r_{\rm np} =
0.15$ fm  in the  $^{208}$Pb nucleus.  {\label{fig11} }}
\end{figure}

We now consider our results for the tidal polarizability parameter
$\lambda$ defined as,
\begin{equation}
Q_{ij}=-\lambda {\cal E}_{ij},
\end{equation}
where, $Q_{ij}$ is the induced quadrupole moment of a star in binary due to
the static external tidal field of the companion star ${\cal E}_{ij}$. The
parameter $\lambda$ can be expressed in terms of the dimensionless
quadrupolar tidal Love  number $k_2$ as,
\begin{equation}
\lambda = \frac{2}{3G}k_2R^5,
\end{equation}
where, $R$ is the radius of a isolated neutron star,
i.e.,  long before merger.  The value of $k_2$ depends on
the stellar structure and can be calculated by  following
the procedure used in Refs. \cite{Flanagan08,Hinderer08,Hinderer10}. The
values of $\lambda$ for the neutron stars with masses $\sim 1M_\odot$
are sensitive to the behaviour of the symmetry energy  at supra-nuclear
densities \cite{Fattoyev13b}. In Fig. \ref{fig11}, we plot the values
of $\lambda$ as a function of neutron star mass obtained for different
parameterizations for the  $F_\rho$ and $F_{2\rho}$ families. The value
of $\Delta r_{\rm np}$ is equal to 0.15 fm for all of these cases.
The differences in the tidal polarizability at  low mass neutron star for
the two different families is very small. But the difference increases as
the mass increases due to different high density behaviour of the symmetry
energy for different families of models.  The values of $\lambda$ for
the neutron star with canonical mass vary over a wide range of $2.7\times
10^{36}$ to $4.3\times 10^{36}$ cm$^2$gs$^2$.  The value of $\lambda$ at
$1.4M_\odot$ obtained  for the $F_\rho$ family is about 1.5 times larger
than that for the $F_{2\rho}$ family.  The inclusion of $\delta$ mesons
slightly lowers  the value of tidal polarizability of neutron star with
mass 1.4$M_\odot$.  On passing, it may be remarked that the differences
in $\lambda$ across the two different families of the models are larger
than the uncertainties expected in its measurement by the advanced
LIGO-Vergo detector and Einstein Telescope \cite{Hinderer10,Damour12}.

\begin{table}
\caption{\label{tab3} Properties of neutron stars and the neutron-skin
thickness in the $^{208}$Pb nucleus obtained  for SET1 -SET4 parameters.
The values of tidal polarizability parameter $\lambda_{1.4}$, listed in the last row, 
correspond to the  neutron star with mass $1.4M_\odot$.
}
\begin{tabular}{|c|cccc|}
\hline
\multicolumn{1}{|c}{}&
\multicolumn{2}{|c}{$F_\rho$}&
\multicolumn{2}{c|}{$F_{2\rho}$}\\
\cline{2-5}
Properties&  SET1& SET2& SET3 &SET4 \\
\hline
$\Delta r_{\rm np}$ (fm)&  0.22& 0.15& 0.22&0.15\\
 $\rho_{\scriptscriptstyle{DU}}$ (fm)$^{-3}$   &  0.297  &     0.282& 0.401&0.505   \\    
 $\rho_t$(fm)$^{-3}$    &      0.058 &   0.069 &0.073 &0.107 \\
 $P_t$   (MeV fm$^{-3}$)     &      0.222 &   0.107 & 0.474 &0.509  \\
 $R_{1.4}$(km)     &       13.08 &   12.96  & 13.00 &12.37   \\
  $R_{\rm max}$(km)       & 11.40   & 11.68 &11.29  &11.28 \\
 $M_{\scriptscriptstyle{DU}} (M_\odot)$      & 1.01    &  1.09 & 1.33   &1.69 \\
$\lambda_{1.4}$ (10$^{36}$ cm$^2$g s$^2$) & 3.41& 4.33&2.87&2.88\\
 \hline
\end{tabular}
\end{table}
Finally, we have collected in Table \ref{tab3} the results for
the various properties of the neutron stars obtained for  a few
representative cases corresponding to the $F_\rho$ and $F_{2\rho}$
families. The values of neutron-skin thickness for the $^{208}$Pb nucleus  are
also listed. The comparison of the results obtained for the SET1 with those for  SET3
or SET2 with SET4 readily gives a crude estimate about the variation
in the properties of the neutron stars across the different families
of models at a fixed neutron-skin thickness. Similarly, the idea
about the effects of $\delta$ mesons within the same family can be
obtained by comparing the results for SET1 with SET2 or SET3 with SET4.
It may be easily verified from Table \ref{tab3} and Figs. \ref{fig8} -
\ref{fig11} that the values of core-crust transition density $\rho_t,
R_{1.4}$, $M_{\scriptscriptstyle{DU}}$ and the tidal
polarizability parameter $\lambda$  for both the families of the models can
differ significantly at a fixed value of $\Delta r_{\rm np}$.  Thus, instead
of the $\sigma - \rho$ and $\omega-\rho$ cross-couplings as included
separately in the different families of the models,  a linear combinations
of these cross-couplings in a single model would allow one to adjust
the properties of the neutron stars over a wide range at a fixed value of
the neutron skin thickness in a heavy nucleus, like, $^{208}$Pb.
Furthermore, the presence of $\delta$ mesons enable ones to obtain 
the models with smaller value of neutron-skin thickness as can be seen
from Fig. \ref{fig3}.

\section{Conclusions}

We have studied the differences in the various properties of the neutron
stars arising mainly due to the uncertainties in the density content of
the nuclear symmetry energy in the extended RMF model. With this aim,
two different families of the   extended RMF model, namely, $F_\rho$ and
$F_{2\rho}$ are obtained.  The $F_{\rho}$ family includes $\sigma-\rho$
cross-coupling, while, the $F_{2\rho}$ family includes $\omega-\rho$
cross-coupling. Both the families of models include the contributions
from the $\delta$ meson in addition to several linear and non-linear
interaction terms already present in the commonly used RMF models. The
several parameterizations for each of the families of the models are so
obtained that they yield wide  variations for the neutron-skin thickness
$\Delta r_{\rm np}$ in  the $^{208}$Pb nucleus without affecting much
the other bulk properties of the nuclei.  The inclusion of $\delta$ meson
produces required splitting  in the effective mass for the neutrons and
protons and also enables us to obtain smaller neutron-skin thickness.

We compare the various properties of neutron stars obtained for the
$F_{\rho}$ and $F_{2\rho}$ families.  The properties of neutron stars
considered are  the core-crust transition density, radius, red shift,
tidal polarizability parameter and threshold  mass required for the
enhanced cooling through direct Urca process. Most of these properties
of the neutron stars at a fixed $\Delta r_{\rm np}$ are  noticeably
different for two different families of the models. These differences
are pronounced at smaller  values of $\Delta r_{\rm np}$ which can be
attributed to the differences in the density dependence of the
symmetry energy  resulting due to different cross-coupling terms.
For $\Delta r_{\rm np} = 0.15 {\rm fm}$ in the $^{208}$Pb nucleus,
consistent with the current experimental data on dipole polarizability ,
the red-shift and the radius of neutron stars with  mass $1.4M_\odot$
differs  by about $10\%$ for the two families of models. Such differences
are quite significant ($\sim 40\%$) for the tidal polarizability parameter
for the neutron stars with mass $1.4M_{\odot}$ and the threshold mass
required for the direct Urca process to occur in the neutron stars. The
values of the core-crust transition density also differs reasonably
across the different families of the models.  We may thus say that
the simultaneous inclusion of the $\sigma-\rho$ and $\omega-\rho$
cross-couplings in the extended RMF model would enhance its flexibility
to accommodate the variations in the properties of the neutron stars at
a given neutron-skin thickness.

\begin{acknowledgments}
A.S. acknowledges the support given by Universitas Indonesia and would
like to thank SINP for the hospitality during his visit  where this
work was initiated.  The authors gratefully acknowledge the assistance
of Tanuja Agrawal in the preparation of the manuscript.
\end{acknowledgments}


\end{document}